\documentclass[conference]{IEEEtran}
\IEEEoverridecommandlockouts

\usepackage{cite}
\usepackage{amsmath,amssymb,amsfonts}
\usepackage{algorithmic}
\usepackage{hyperref}
\usepackage{graphicx}
\usepackage{multirow}
\usepackage{bbding}
\usepackage{textcomp}
\usepackage{makecell}
\usepackage{array}
\usepackage{booktabs}
\usepackage{pgfmath}
\usepackage{utfsym}
\usepackage{fontawesome}
\usepackage{float}
\usepackage{url}
\usepackage[numbers,sort&compress]{natbib}

\usepackage{tabularx}
\usepackage{pifont}
\usepackage{subcaption}
\usepackage{comment}

\usepackage{pifont} % include some cool symbols
\definecolor{bsRed}{rgb}{0.95, 0.0, 0.0}

\def\BibTeX{{\rm B\kern-.05em{\sc i\kern-.025em b}\kern-.08em
    T\kern-.1667em\lower.7ex\hbox{E}\kern-.125emX}}
\begin{document}

\title{ SLM-S2ST: A multimodal language model for direct speech-to-speech translation
% Phi-Omni-ST: Enabling Speech-aware Language Models with Speech-out Capacities
% Phi4-Omni for speech-to-speech translation
% Enabling Phi-4-Multimodal for speech-to-speech translation
\thanks{* Equal contribution.}
}

% \author{Anonymous Author}

\author{
Yuxuan Hu*, Haibin Wu*, Ruchao Fan, Xiaofei Wang, Heng Lu, Yao Qian, Jinyu Li
\\
\textit{Microsoft, USA}
}

\maketitle

\begin{abstract}
Speech-aware language models (LMs) have demonstrated capabilities in understanding spoken language while generating text-based responses. However, enabling them to produce speech output efficiently and effectively remains a challenge.  In this paper, we present SLM-S2ST, a multimodal LM for direct speech-to-speech translation (S2ST), built on the open-source Phi4-MM model. SLM-S2ST extends its predecessor by generating translated speech using an audio transformer head that predicts audio tokens with a delay relative to text tokens, followed by a streaming vocoder for waveform synthesis. Our experimental results on the CVSS-C dataset demonstrate SLM-S2ST's superior performance, significantly surpassing existing baseline models trained on the same dataset. Furthermore, when we scale up the training data and the model size, SLM-S2ST reaches on-par performance with the current SOTA model.
\end{abstract}


\section{Introduction}
\label{sec:intro}
% To be revised
Enabling fluent and seamless communication across languages through speech is a long-standing goal in spoken language technology. Speech-to-speech translation addresses this challenge by directly converting spoken utterances from a source language into speech output in a target language. Traditional systems typically decompose the task into three separate stages—speech recognition, machine translation, and speech synthesis. While this cascaded method is effective to some extent, it is often hindered by error propagation, rigid interfaces between components, and high latency, making real-time, high-quality translation difficult to achieve.

Speech language models (Speech LMs)~\cite{arora2025landscape,defossez2024moshi,ji2024wavchat,wu2024towards,peng2024survey} have garnered increasing attention from both academia and industry, especially following the release of GPT-4o~\cite{gpt4o}. 
A prevailing class of speech language models, known as speech-aware LMs, focuses on understanding speech and generating text responses. These models have been extensively studied, with high-performing systems emerging rapidly in recent years. 
Speech-aware LMs~\cite{tang2023salmonn,gong2023listen,chu2023qwen,ghosh2025audio,phi4mini} typically integrate pre-trained speech encoders with text-based LLMs to preserve the instruction-following (IF) capabilities of the language model while enabling reasoning over spoken input~\cite{phi4mini,chu2023qwen}. 
They are designed to model the conditional distribution $p(text|speech,text)$, producing appropriate text outputs based on both speech and textual instructions.
One notable example is Phi4-MM~\cite{phi4mini}, which demonstrates excellent speech understanding capabilities and achieves state-of-the-art performance on tasks such as automatic speech recognition and speech-to-text translation.
A great number of speech-aware language models are open-source, making them strong starting points for building speech-to-speech conversational systems. If such models can be adapted from speech-to-text interaction to speech-to-speech interaction using affordable resources, it would be a game changer—opening up speech language model development to the broader research community, not just limited to large companies with vast computational resources.

However, how to efficiently and effectively extend these speech-aware LMs to speak remains an open question. 
This requires modeling a richer distribution $p(speech,text|speech,text)$, instead of $p(text|speech,text)$ as in speech-aware LMs, with the capacity to generate coherent spoken outputs alongside text, with minimal required resources.
After the extension, these models can perceive speech input and then directly generate speech output within a unified, end-to-end framework—without explicitly converting speech to text as intermediate steps. 
As a result, by end-to-end modeling speech, these models offer the potential to alleviate error propagation, reduce latency, and better preserve paralinguistic information, making them well-suited for the next generation of speech translation systems.

\begin{figure*}[htp]
    \centering
    {\includegraphics[width=0.9\textwidth]{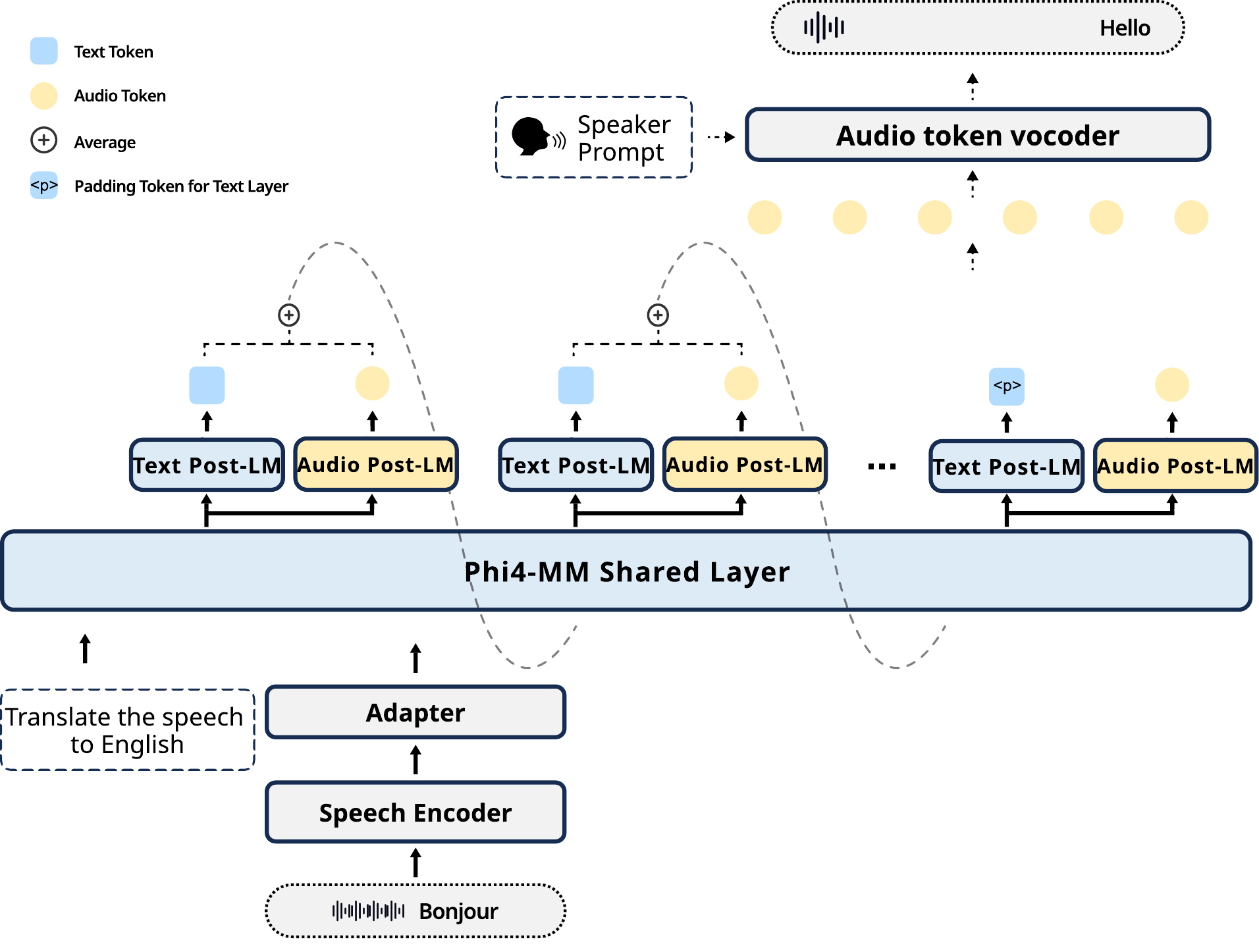}}
    \caption{The proposed framework for end-to-end speech-to-speech translation. The Phi4-MM model includes Phi4-MM Shared Layer and Text Post-LM. The Audio Post-LM is initialized with the Text Post-LM, and is trainable.}
    \label{figure:framework}
\end{figure*}

In this paper, we present SLM-S2ST, a multimodal \textbf{S}peech-aware \textbf{L}arge \textbf{L}anguage \textbf{M}odel for \textbf{S}peech to \textbf{S}peech \textbf{T}ranslation, built on Phi4-MM model.
This proposed model integrates both speech understanding and speech generation within a single model. 
Specifically, we employ an audio transformer head to predict audio tokens with a delay relative to the text tokens generated by the language model. The generated speech tokens are then passed through a streaming token-to-mel-spectrogram module, followed by a causal vocoder to synthesize the final speech waveform.
Our main contributions are as follows.
\begin{itemize}
    \item We propose a simple yet effective approach to extend a well-pretrained speech-aware multimodal large language model for speech-to-speech translation (S2ST). By training SLM-S2ST solely on the CVSS-C dataset, we achieve state-of-the-art performance—surpassing other models trained on the same data by a substantial margin (e.g., improving BLEU from 20.93 to 33.96 on DE-EN S2ST). To the best of our knowledge, this is the first work to adapt and extend Speech-aware LMs for the S2ST task.
    
    \item We further scale up our approach by increasing the amount of training data to around 11,000 hours and increasing the model size from 3.8B to 7B. This results in substantial performance gains, performing on-par with open-source state-of-the-art models. % trained on larger scale of data.
    
    \item Through extensive experiments, we demonstrate that our method achieves strong alignment between output text and speech, enabling near-lossless transfer of capabilities from speech-to-text to speech-to-speech translation. The proposed method is not limited to S2ST but is also applicable to other speech-to-speech tasks. % The proposed method is not limited for S2ST, but other speech-to-speech tasks.
\end{itemize}
\section{Method}
\label{sec:method}

The main framework is shown in Figure~\ref{figure:framework}.
To enable speech understanding, we incorporate a speech encoder and an adapter module before the language model, to take speech as inputs and output representations that could be directly fed into the language model, following the same architecture used in Phi4-MM. 
The Phi4-MM shared layer (the main LM backbone) takes both the text instructions (``Translate the speech to English'') and the speech embeddings as inputs, and output hidden embeddings for the text post-LM and audio post-LM to jointly model and generate text and speech tokens. The decoded text tokens produce the text translation, while the generated speech tokens are passed through a streaming speech token vocoder to synthesize the final translated waveform. 

\subsection{Speech Understanding}
Our speech encoder and adapter follow the same architecture as in Phi4-MM. Both components are initialized from Phi4-MM and kept frozen during training. 
The speech encoder consists of three convolutional layers followed by 24 Conformer blocks~\cite{conformer20}, each with 1,024 attention dimensions, 1,536 feed-forward dimensions, and 16 attention heads. The convolutional layers perform temporal downsampling, producing embeddings at a 12.5 Hz frame rate. A two-layer MLP speech adapter then maps these speech representations into the same embedding space and dimensionality as the text tokens.
The Phi4-MM Shared Layer and Text Post-LM in Figure~\ref{figure:framework}, composite the original Phi4-MM language model, using several layers of Transformers.

The speech encoder, adapter, Phi4-MM shared layers, and text post-LM layer in Figure~\ref{figure:framework}, have been trained on multilingual speech data as demonstrated in \cite{phi4mini}, giving the model strong capabilities in listening and understanding multilingual speech. 
We take their developed open-source Phi4-MM models~\cite{phi4mini} as our initializations.
Our goal is to leverage their multi-lingual speech understanding capacity and, with minimal adaptation using small-scale data, enable the model to extend from speech-to-text translation to speech-to-speech translation.

\subsection{Jointly text-speech decoding}
We adopt the simultaneous text-speech joint decoding paradigm~\cite{ding2025kimi}, where text and audio tokens are decoded together in a single forward pass of the language model, as illustrated in Figure~\ref{figure:framework}. 
Unlike~\cite{ding2025kimi}, which trains the entire model from scratch using massive computational resources, millions of hours of data and complex different-stage training, our approach focuses on enabling speech generation through modest resources and lightweight adaptation with only one-stage fine-tuning. 
We show that strong speech-to-speech translation performance can be achieved using only a few hundred hours of open-source data~\cite{jia2022cvss}, with our base model also derived from an open-source release~\cite{phi4mini}. 
This makes our method much easier to reproduce and is accessible to the research community, significantly lowering the barrier to entry and promoting broader participation in speech language model development beyond large industry players.

In our approach, as illustrated in Figure~\ref{figure:framework}, both text and audio tokens are generated simultaneously in a single forward pass of the main language model. 
The text post-LM consists of the last few layers from the Phi4-MM model, and we randomly initialize the audio post-LM using the same architecture as the text post-LM. 
After the shared Phi4-MM layers produce hidden representations, the text post-LM uses these representations to predict text tokens, while the audio post-LM simultaneously predicts audio tokens using the same hidden states. 
These predicted tokens are then averaged in the hidden space and passed back into the language model for the next forward step. 
As shown in prior work~\cite{pGSLM}, introducing a delay for audio token generation allows the model to utilize future text tokens as context, thereby improving the quality of audio token prediction through look-ahead.
We also include a delay of 3 for the audio tokens.

The speech tokens in Figure~\ref{figure:framework} are extracted using a pretrained speech tokenizer based on finite scalar quantization (FSQ)\cite{mentzer2024finite}. The FSQ module is applied to an ASR encoder initialized from the Whisper model\cite{radford2023robust}, where intermediate representations are projected into a lower-dimensional space and then discretized via a rounding operation. The tokenizer is trained using an ASR loss. In our work, we directly adopt the pretrained speech tokenizer provided by CosyVoice 2~\cite{du2024cosyvoice}, without any additional fine-tuning.

\subsection{Streaming Speech Decoder}
The speech decoder is performed in a streaming way. The speech tokens generated by the language model are grouped into chunks of 10, and each chunk, along with a target speaker prompt, is passed to a causal flow matching model~\cite{lipman2023flow} to synthesize mel-spectrograms. Each chunk of tokens yields a segment of mel-spectrogram, enabling streaming synthesis. The resulting spectrograms are then converted into waveforms using a HiFi-GAN vocoder~\cite{hifigan}. Both the flow matching model and the vocoder are adopted directly from the pretrained CosyVoice 2\footnote{https://github.com/FunAudioLLM/CosyVoice} without further modification.

\section{Experimental setup}
\label{subsec:exp_setup}
\begin{table}[ht]
\centering
\setlength{\tabcolsep}{4pt}
\renewcommand{\arraystretch}{1.15}
\caption{7 Languages of training data. Both source and target speech durations are included in the total.}
\label{tab:dataset_stats}
\begin{tabular}{cccc}
\toprule
\textbf{Languages} & \textbf{CVSS-C (hrs)} & \textbf{CVSS-M (hrs)} & \textbf{In-house data (hrs)} \\
\midrule
DE & 275 & 287 & 1,448 \\
ES & 165 & 121 & 1,373 \\
FR & 368 & 382 & 1,434 \\
IT & 68  & 71  & 1,352 \\
JA & 2   & 2   & 1,573 \\ 
PT & 15  & 15  & 1,386 \\
ZH & 12  & 12  & 1,793 \\
Total & 905 & 940 & 10,359 \\
% \textbf{Languages} & \textbf{CVSS-M (hrs)} & \textbf{In-house data (hrs)} \\
% \midrule
% DE & 287 & 1,448 \\
% ES & 121 & 1,373 \\
% FR & 382 & 1,434 \\
% IT & 71  & 1,352 \\
% JA & 2   & 1,573 \\ 
% PT & 15  & 1,386 \\
% ZH & 12  & 1,793 \\
% Total & 940 & 10,359 \\
\bottomrule
\end{tabular}
\end{table}
% DE: 17.41, ES: 18.12, FR: 18.3, IT: 19.92
% \input{tables/readout}
\subsubsection{Training Dataset} 

For the CVSS-M dataset, we leveraged a speech-to-speech translation dataset based on CoVoST2~\cite{wang2021covost2} by using the original source-language speech paired with target-language text. We selected seven target languages from the dataset: German (DE), Spanish (ES), French (FR), Italian (IT), Japanese (JA), Portuguese (PT), and Chinese (ZH). 
A zero-shot TTS model was then employed to synthesize multi-speaker target-language speech. Finally, we used S3tokenizer\footnote{https://github.com/xingchensong/S3Tokenizer} to convert the speech into semantic tokens. The resulting dataset, referred to as CVSS-M, contains approximately 940 hours of data as shown in Table~\ref{tab:dataset_stats}.

In addition to CVSS-M, we include the CVSS-C dataset~\cite{jia2022cvss} for comparison. While both are built on CoVoST2 and share the same source speech, CVSS-C differs in that its target speech is synthesized using a single-speaker TTS system, resulting in more uniform acoustic characteristics. To maintain consistency with CVSS-M and avoid added complexity, we use the non-normalized (non-TN) version of the target text in CVSS-C. This setup ensures a fair comparison by isolating the impact of acoustic variability, as CVSS-C poses a lower learning burden than the more acoustically diverse CVSS-M. The statistics are shown in Table~\ref{tab:dataset_stats}

To further scale training, we incorporated an additional 10,000 hours of data derived from the Phi4-MM speech translation task. In this extended dataset, the target-language text was converted into speech using the same zero-shot TTS pipeline as CVSS-M. This augmentation significantly improves performance. More details are shown in Table~\ref{tab:dataset_stats}.

\subsubsection{Training details} 
For the audio post-LM, we follow a similar design to~\cite{ding2025kimi}, using the last six Transformer decoder layers from the open-source Phi4-MM architecture\footnote{https://huggingface.co/microsoft/Phi-4-multimodal-instruct}, with identical configurations. 
This module is randomly initialized and requires no additional pretraining. 
We incorporate a Low-Rank Adaptation (LoRA)\cite{hu2022lora} module with rank 320, applied only to the original LLM decoder layers—specifically the Phi4-MM shared layer and the text post-LM, as shown in Figure~\ref{figure:framework}. During training, we freeze the base LLM parameters and update only the audio post-LM, LoRA modules, the audio LM head, and a linear speech-out prediction layer.
For the Phi4-MM base models, we consider two configurations: the 4B model as described in \cite{phi4mini}, and a scaled-up 7B version.
% The model is trained with a learning rate of 1e-4, using 32 H100 GPUs. Training takes approximately 16 hours to complete.

\subsubsection{Evaluation data}
For the testing data, we adopt CoVoST2~\cite{wang2021covost2}, FLEURS~\cite{conneau2023fleurs} and CVSS~\cite{jia2022cvss} test sets.
\begin{itemize}
    \item CoVoST2 is a large-scale multilingual speech translation dataset. It supports both X-to-English (X-En) and English-to-X (En-X) directions. In our experiments, we evaluate performance on the X-En translation task using the official test sets for the supported languages
    \item For FLEURS, we use the same audio samples as those used in the ASR evaluation in~\cite{phi4mini}, but replace the ASR references with their corresponding translations. As with CoVoST2, we focus on X-En directions.
    \item CVSS is a massively multilingual-to-English speech-to-speech translation corpus, it is derived from Common Voice speech corpus and the CoVoST2 speech-to-text translation corpus. The inputs of CVSS test set are exactly the same as in CoVoST2, but the label text has been normalized, making it directly suitable for computing ASR-BLEU scores.
\end{itemize}

\subsubsection{Evaluation metrics}
Our model is capable of generating both text and audio translation outputs simultaneously, so we evaluate these two modalities separately. 
\begin{itemize}
    \item For Speech-to-Text Translation (S2TT), we compute BLEU scores on CoVoST2~\cite{wang2021covost2} and FLEURS~\cite{conneau2023fleurs} test sets, comparing the model’s text predictions against the provided ground-truth references. 
    \item For Speech-to-Speech Translation (S2ST), we use the ASR-BLEU metric: the generated speech is first transcribed using Whisper-Large-V3~\cite{radford2023robust}\footnote{https://huggingface.co/openai/whisper-large-v3}, and both the transcriptions and the reference texts from CVSS~\cite{jia2022cvss} are normalized using the official Whisper text normalizer. The normalized transcriptions and the text references are used to calculate the BLEU scores. Our evaluation protocol follows that of SeamlessM4T~\cite{barrault2023seamless} to ensure consistency and comparability with prior works.
    \item We measure the word error rate (WER) between the transcribed speech and the model-generated text using Whisper-large-v3. This metric evaluates the alignment between the language model's generated speech and its corresponding text output.
\end{itemize}

\subsubsection{Baselines}

For models trained solely on the CVSS-C corpus~\cite{jia2022cvss}\footnote{Note that the settings are not fully controlled; some papers report using CVSS-C but do not specify whether all languages are included.}, we compare five representative S2ST systems:
\begin{itemize}
    \item S2UT~\cite{lee2022direct}, a one-pass non-autoregressive speech-to-unit translation model;
    \item Translatotron 2~\cite{jia2022translatotron}, an end-to-end, one-pass speech-to-speech translation system;
    \item DASpeech~\cite{fang2023daspeech}, which employs a two-pass non-autoregressive architecture for both the speech-to-text and text-to-unit stages;
    \item UnitY~\cite{inaguma2023unity}, which uses a fully autoregressive two-pass processing pipeline;
    \item StreamSpeech~\cite{zhang2024streamspeech}, with a two-stage process, which introduces an autoregressive first pass for reordering and contextual modeling, followed by a non-autoregressive second pass for unit decoding.
\end{itemize}

Another line of baselines includes state-of-the-art systems pretrained on tens of thousands of hours of data:
\begin{itemize}
    \item Seamless M4T v2 Large~\cite{barrault2023seamless} offers dynamic modality switching for robust end-to-end speech-to-speech translation and represents the current SOTA, trained on millions of hours of multilingual data;
    \item Whisper V3~\cite{radford2023robust} is trained on ASR, speech translation (ST), speaker diarization, and other tasks. This model can support speech-to-text translation;
    \item Qwen2-audio~\cite{chu2024qwen2} is a speech-aware language model trained on large-scale data for speech understanding, including the spech-to-text translation;
    \item Qwen2.5 Omni~\cite{xu2025qwen2}\footnote{Qwen2.5 Omni claims to support speech-to-text translation and speech generation capabilities, but does not explicitly claim the ability of speech-to-speech translation. From our results, it appears that including speech-to-speech translation data into training is still necessary to enable their S2ST capability.} builds on large-scale audio–language pretraining, supporting both speech understanding and generation. They leverage a Thinker-Talker mode, where the thinker (LM) is responsible for generating text, and the talker takes the text and hidden representations from the LM to generate the speech tokens. We include its performance in our evaluations;
    \item Gemini 2.0 Flash is a commercial model known for its good speech understanding capacities;
    \item GPT-4o~\cite{gpt4o} is an enterprise-level multimodal language model with strong speech-to-speech capacities, developed by OpenAI.
\end{itemize}

\section{Experimental results}

\begin{table}[ht]
\centering
\caption{Models trained by only CVSS data}
\label{tab:cvss_results}
\begin{tabular}{lccc}
\toprule
\textbf{Model} & \multicolumn{1}{c}{\textbf{FR-EN}} & \multicolumn{1}{c}{\textbf{ES-EN}} & \multicolumn{1}{c}{\textbf{DE-EN}} \\
                   & ASR-BLEU $\uparrow$ & ASR-BLEU $\uparrow$ & ASR-BLEU $\uparrow$ \\
\midrule
S2UT~\cite{lee2022direct}             & 22.23 & 18.53 & 2.99 \\
Translatotron 2~\cite{jia2022translatotron}  & 26.07 & 22.93 & 16.91 \\
DaSpeech~\cite{fang2023daspeech}         & 25.03 & 21.37 & 16.14 \\
UnitY~\cite{inaguma2023unity}            & 27.77 & 24.95 & 18.74 \\
StreamSpeech~\cite{zhang2024streamspeech}     & 28.45 & 27.25 & 20.93 \\
\textbf{Ours 4B (CVSS-C)}          & \textbf{35.93} & \textbf{39.67} & \textbf{35.54} \\
\bottomrule
\end{tabular}
\end{table}
\begin{table*}[ht]
\centering
\footnotesize
\setlength{\tabcolsep}{3pt}
\renewcommand{\arraystretch}{1.3}
\caption{Performance on CoVoST2 X-EN. S2ST results are measured on the CVSS test set. Each cell reports the S2T / S2S score (higher is better); ‘--’ denotes that the S2S score is unavailable. For each language, the highest S2T score is shown in bold and the second-highest is underlined.}
\begin{tabular}{llcccccccc}
\toprule
\textbf{Category} & \textbf{Model} & \textbf{DE $\uparrow$} & \textbf{ES $\uparrow$} & \textbf{FR $\uparrow$} & \textbf{IT $\uparrow$} & \textbf{JA $\uparrow$} & \textbf{PT $\uparrow$} & \textbf{ZH $\uparrow$} & \textbf{AVG $\uparrow$} \\
\midrule
\multirow{6}{*}{\textbf{S2T}} 

& Base model-4B           
& \underline{39.81} / -- 
& \underline{43.60} / -- 
& \underline{42.24} / -- 
& \underline{41.42} / -- 
& \underline{30.54} / -- 
& \underline{55.28} / -- 
& 22.39 / -- 
& \underline{39.33} / -- \\

& Base model-7B      
& \textbf{41.27} / -- 
& \textbf{44.77} / -- 
& \textbf{43.39} / -- 
& \textbf{42.73} / -- 
& \textbf{31.79} / -- 
& \textbf{56.43} / -- 
& \textbf{24.25} / -- 
& \textbf{40.65} / -- \\

& Whisper V3        
& 34.17 / -- 
& 39.21 / -- 
& 35.43 / -- 
& 35.82 / -- 
& 23.59 / -- 
& 50.22 / -- 
& 14.36 / -- 
& 33.26 / -- \\

& Qwen2-audio       
& 34.99 / -- 
& 39.91 / -- 
& 38.31 / -- 
& 36.35 / -- 
& 22.98 / -- 
& 47.79 / -- 
& \underline{23.27} / -- 
& 34.80 / -- \\

& Gemini-2.0-Flash  
& 38.34 / -- 
& 41.74 / -- 
& 38.96 / -- 
& 37.76 / -- 
& 28.04 / -- 
& 50.81 / -- 
& 20.69 / -- 
& 36.62 / -- \\

& GPT-4o            
& 39.29 / -- 
& 41.49 / -- 
& 38.56 / -- 
& 37.33 / -- 
& 30.46 / -- 
& 50.60 / -- 
& 21.93 / -- 
& 37.09 / -- \\

\midrule
\multirow{6}{*}{\textbf{S2S}} 
& Qwen2.5 Omni      & \textbf{39.91} / 10.50 & \textbf{42.92} / 9.84 & \underline{41.51} / 9.97 & \underline{39.68} / 9.66 & 23.98 / 11.59 & \underline{51.01} / 17.05 & \textbf{26.19} / 5.04 & \textbf{37.89} / 10.52 \\

& SeamlessM4T-V2    & \underline{39.90} / \textbf{41.90} & \underline{42.90} / \textbf{45.34} & \textbf{42.18} / \textbf{55.29} & \textbf{39.85} / 25.17 & 22.18 / 9.93 & \textbf{53.82} / \textbf{57.45} & 21.92 / \textbf{22.82} & \underline{37.54} / \underline{36.84} \\

& Ours 4B (CVSS-M)    & 35.64 / 33.96 & 39.20 / 37.88 & 36.07 / 33.70 & 36.29 / 34.47 & 24.80 / 23.74 & 49.44 / 49.94 & 20.62 / 17.94 & 34.58 / 33.09 \\

& Ours 4B (CVSS-C)    & 35.55 / 35.84 & 39.39 / 39.67 & 36.69 / 35.93 & 36.05 / 36.45 & 25.21 / 24.62 & 49.27 / 50.81 & 20.89 / 19.70 & 34.72 / 34.72 \\

& Ours 4B (In+CVSS-M)           & 36.13 / 37.98 & 37.45 / 39.62 & 34.64 / 35.48 & 34.42 / \underline{36.51} & \underline{28.14} / \underline{28.45} & 48.58 / 51.21 & 20.65 / 20.66 & 34.29 / 35.70 \\

& Ours 7B (In+CVSS-M)          & 38.08 / \underline{39.71} & 39.51 / \underline{41.42} & 36.79 / \underline{37.27} & 36.49 / \textbf{38.36} & \textbf{30.38} / \textbf{30.98} & 50.40 / \underline{52.78} & \underline{22.48} / \underline{22.39} & 36.30 / \textbf{37.56} \\

\bottomrule
\end{tabular}
\label{tab:covost-results}
\end{table*}

\begin{table*}[ht]
\centering
\footnotesize
\setlength{\tabcolsep}{3pt}
\renewcommand{\arraystretch}{1.3}
\caption{Performance on Fleurs X-EN. Each cell reports the S2T / S2S score (higher is better); ‘–’ denotes that the S2S score is unavailable. For each language, the highest S2T score is shown in bold and the second-highest is underlined.}
\begin{tabular}{lccccccccc}
\toprule
\textbf{Category} & \textbf{Model} & \textbf{DE $\uparrow$} & \textbf{ES $\uparrow$} & \textbf{FR $\uparrow$} & \textbf{IT $\uparrow$} & \textbf{JA $\uparrow$} & \textbf{PT $\uparrow$} & \textbf{ZH $\uparrow$} & \textbf{AVG $\uparrow$} \\
\midrule
\multirow{6}{*}{\textbf{S2T}} 

& Base model-4B           & 37.71 / -- & 25.33 / -- & 35.10 / -- & 26.06 / -- & 21.62 / -- & 40.80 / -- & 22.37 / -- & 29.86 / -- \\

& Base model-7B        & \underline{39.33} / -- & 26.32 / -- & \underline{36.31} / -- & \underline{27.53} / -- & 23.05 / -- & \underline{42.95} / -- & \textbf{24.37} / -- & \underline{31.41} / -- \\

& WhisperV3         & 33.49 / -- & 22.68 / -- & 30.98 / -- & 23.00 / -- & 16.63 / -- & 37.50 / -- & 16.07 / -- & 25.76 / -- \\

& Qwen2-audio       & 32.88 / -- & 22.40 / -- & 30.82 / -- & 22.12 / -- & 4.49 / -- & 35.38 / -- & 17.95 / -- & 23.72 / -- \\

& Gemini-2.0-Flash  & 38.48 / -- & \underline{26.51} / -- & 35.18 / -- & 25.02 / -- & \underline{23.89} / -- & 41.51 / -- & \underline{24.27} / -- & 30.69 / -- \\

& GPT-4o            & \textbf{41.03} / -- & \textbf{29.10} / -- & \textbf{37.98} / -- & \textbf{28.51} / -- & \textbf{24.17} / -- & \textbf{43.33} / -- & 24.12 / -- & \textbf{32.61} / -- \\

\midrule
\multirow{6}{*}{\textbf{S2S}} 

& Qwen2.5 Omni      & \textbf{39.61} / 6.43 & \textbf{27.58} / 4.69 & \textbf{36.98} / 6.88 & \textbf{28.13} / 4.68 & \underline{20.60} / 5.11 & \textbf{43.38} / 7.66 & \textbf{28.28} / 6.26 & \textbf{32.08} / 5.94 \\

& SeamlessM4T-V2    & 36.80 / \textbf{41.12} & \underline{25.67} / \textbf{28.00} & \underline{33.78} / \textbf{37.31} & \underline{26.80} / \textbf{29.59} & 18.63 / \underline{21.51} & 37.61 / \textbf{43.62} & \underline{22.78} / \textbf{23.95} & 28.87 / \textbf{32.14} \\
% & Ours    & &  &  & &   &  &   & \\
& Ours 4B (CVSS-M)    & 32.57 / 25.29 & 22.61 / 16.61 & 31.44 / 25.17 & 23.01 / 18.29 & 17.18 / 12.92 & 34.89 / 28.74 & 18.42 / 14.65 & 25.73 / 20.24 \\

& Ours 4B (CVSS-C)    & 32.14 / 29.26 & 21.84 / 18.55 & 30.18 / 27.12 & 22.58 / 19.96 & 17.23 / 15.55 & 35.01 / 31.87 & 18.21 / 16.13 & 25.31 / 22.63 \\

& Ours 4B (In+CVSS-M)          & 35.89 / 37.83 & 23.27 / \underline{24.94} & 33.09 / 35.26 & 25.13 / 26.53 & 19.11 / 19.94 & 38.98 / 41.42 & 20.91 / 21.48 & 28.05 / 29.69 \\

& Ours 7B (In+CVSS-M)          & \underline{37.89} / \underline{39.55} & 24.01 / 24.93 & 33.67 / \underline{35.71} & 26.72 / \underline{27.60} & \textbf{21.26} / \textbf{21.67} & \underline{40.78} / \underline{43.59} & 22.60 / \underline{22.72} & \underline{29.56} / \underline{30.96} \\

\bottomrule
\end{tabular}
\label{tab:fleurs-results}
\end{table*}

% & Phi4-ST 4B S1 & 33.00 / 34.96 & 21.86 / 23.51 & 27.85 / 30.17 & 20.96 / 22.96 & 16.67 / 16.86 & 35.65 / 38.74 & 19.98 / 20.49 & 25.14 / 26.81 \\
% & Phi4-ST ESI 4B & 34.79 / 37.01 & 23.35 / 25.02 & 32.25 / 34.89 & 24.89 / 26.84 & 18.45 / 19.02 & 38.36 / 41.63 & 20.41 / 21.07 & 27.50 / 29.35 \\
% & Phi4-ST ESI 7B & 37.67 / 39.79 & 23.88 / 25.26 & 33.61 / 36.12 & 26.44 / 28.58 & 21.21 / 21.93 & 40.09 / 43.12 & 22.68 / 22.95 & 29.37 / 31.11 \\

% & Phi4-ST EP 4B & 35.89 / 37.83 & 23.27 / 24.94 & 33.09 / 35.26 & 25.13 / 26.53 & 19.11 / 19.94 & 38.98 / 41.42 & 20.91 / 21.48 & 28.05 / 29.69 \\
% & Phi4-ST EP 7B & 37.89 / 39.55 & 24.01 / 24.93 & 33.67 / 35.71 & 26.72 / 27.60 & 21.26 / 21.67 & 40.78 / 43.59 & 22.60 / 22.72 & 29.56 / 30.96 \\

\subsection{Model performance trained on CVSS-C data}
The results of models trained solely on the CVSS-C dataset (940 hour in total to include both source and target speech) and evaluated on the CVSS test set are presented in Table~\ref{tab:cvss_results}. We have the following observations:
\begin{itemize}
    \item \textbf{SOTA results:} Our model, surpasses all current baselines trained on the same data by a significant margin—for example, improving BLEU scores on FR-EN from 28.45 to 35.93 (relative 26.3\% improvements), ES-EN from 27.25 to 39.67 (relative 45.6\% improvements), and DE-EN from 20.93 to 35.54 (relative 69.9\% improvements).
    \item As shown in Table~\ref{tab:covost-results}, ``Ours 4B (CVSS-C)'' achieves an average ASR-BLEU score of 34.72, already approaching the performance of ``Ours 4B (in-house)'' trained on 11k hours of data, which scores 35.70 (only 2.7\% relative gap). This demonstrates that our method efficiently transfers speech-to-text translation capabilities to speech-to-speech translation using only small-scale data.
    \item Furthermore, as shown in Table~\ref{tab:covost-results}, our 4B (CVSS-C) model—trained with only 2 hours of Japanese data (Table~\ref{tab:dataset_stats})—achieves an ASR-BLEU score of 23.74. This is a strong result given the extremely limited training data, demonstrating that our training procedure and methodology can yield reasonable performance even with minimal data.
    % This surpasses SeamlessM4T-V2, which was trained on 5.7k hours of Japanese S2ST data, not to mention its overall training data scale. 
    For Chinese (ZH) with 12 hours data and Portuguese (PT)  with 15 hours of data, our model also achieves reasonable results.
\end{itemize}

\subsection{Scale-up the model size and data size}

\begin{table}[ht]
\centering
\setlength{\tabcolsep}{10pt}
\caption{WER (\%) of different models on CoVoST2 and FLEURS datasets.}
\begin{tabular}{ccc}
\toprule
\textbf{Model} & \textbf{CoVoST2 $\downarrow$} & \textbf{FLEURS} $\downarrow$ \\
\midrule
Ours 4B (CVSS-M) & 10.87 & 27.27 \\
Ours 4B (CVSS-C) & 7.80 & 20.62 \\
Ours 4B (In+CVSS-M)       & 4.76 & 2.33  \\
\bottomrule
\label{tab:scaling_wer}
\end{tabular}
\end{table}

We further scale up our approach by increasing the model size from 4B to 7B and augmenting the training data with an additional 10k hours of in-house speech translation data. The main results are presented in Tables~\ref{tab:covost-results} and Table~\ref{tab:fleurs-results}.
The base models, referred to as ``Base model-4B'' and ``Base model-7B'', are used to initialize our systems.
We denote our models as ``Ours [size] ([dataset])''. For example, ``Ours 4B (CVSS-C)'' refers to the model initialized with the 4B base and trained on the CVSS-C dataset. (In+CVSS-M) denotes combining the in-house and CVSS-M data.
We have the below observations:
\begin{itemize}
    \item \textbf{Comparison with current SOTA model:} On the widely used CoVoST2 test set, ``Our 7B (in-house)'' model reaches 37.56 average ASR-BLEU score for speech-to-speech translation, outperforming SeamlessM4T-V2 at 36.84 BLEU. Note that in Italian our system scores 38.36 ASR-BLEU versus SeamlessM4T-V2’s 25.17 ASR-BLEU (-13 ASR-BLEU for SeamlessM4T-V2)
    %, and in Japanese we get 30.98 ASR-BLEU against their 9.93 ASR-BLEU. 
    Our results are achieved with about 11k hours of S2ST data (source + target speech counted together). By contrast, SeamlessM4T-V2 leverages in total about 20k\footnote{SeamlessM4T-V2 does not state whether those hours cover only source speech / target speech or both sides.} hours of data for the seven languages. On Fleurs, a smaller evaluation corpus than CoVoST2, ``Our 7B (in-house)'' model stays on par with SeamlessM4T-V2 in overall speech-to-speech average score. % and gets better scores in Japanese (21.67 vs 21.51).
    \item \textbf{Scaling up the data:} Comparing ``Ours 4B (CVSS-M)'' with ``Ours 4B (In+CVSS-M)'' in Table~\ref{tab:covost-results} and Table~\ref{tab:fleurs-results}, by augmenting CVSS-M with our in-house corpus yields clear improvements: average CoVoST2 S2ST ASR-BLEU rises from 33.09 to 35.7 (+2.61), and Fleurs S2ST ASR-BLEU jumps from 20.24 to 29.69 (+9.45).
    \item \textbf{What a larger model gives us:} As shown in Table~\ref{tab:covost-results} and Table~\ref{tab:fleurs-results}, by enlarging the model from 4B to 7B improves speech-to-speech translation by about 2 AVG ASR-BLEU on CoVoST2 and 1.3 AVG ASR-BLEU on Fleurs, and adds around 1.5 BLEU to speech-to-text on both datasets. These results show that making the model larger with our approach consistently boosts performance and suggests the potential of our approach.
\end{itemize}

We assess speech–text alignment with word error rate (WER) and detailed numbers are listed in Table \ref{tab:scaling_wer}. We have the following observations:
\begin{itemize}
    \item \textbf{CVSS-C vs. CVSS-M:} The model trained on CVSS-C yields a noticeably lower WER than the one trained on CVSS-M (e.g., 7.8\% vs. 10.87\% for CoVoST2 and 20.62\% vs 27.27\% for FLEURS). CVSS-C~\cite{jia2022cvss} uses a single, carefully chosen target speaker, so the acoustic condition is uniform and therefore easier for the model to learn. The same advantage is reflected in higher S2ST BLEU scores in Table \ref{tab:cvss_results} and Table \ref{tab:fleurs-results}, echoing the observations in the original CVSS paper \cite{jia2022cvss}.
    \item \textbf{Adding in-house data:} Comparing ``Ours 4B (CVSS-M)'' with ``Ours 4B (In+CVSS-M)'', by incorporating additional in-house data to the CVSS-M data, the WER drops significantly (e.g., 7.8\% vs. 4.76\% for CoVoST2 and 20.62\% vs 2.23\% for FLEURS), indicating that scaling up training data clearly enhances the alignment of output speech and output text of the LM, and leads to improved output speech quality.
\end{itemize}

\subsection{Discussions}
Our methods have the below advantages:
\begin{itemize}
    \item Extendability to other tasks: While our method is only validated on speech-to-speech translation with strong results, the underlying approach and experience are broadly applicable to other speech-to-speech tasks.
    \item Jointly speech-text modeling: Unlike traditional pipelines that rely on separate ASR, MT, and TTS stages, which would drop information and cumulate errors, our method maintains end-to-end coherence by sharing a unified representation space across modalities.
    \item Based on open-source resources: Our model builds on the open-source Phi4-MM and uses publicly available datasets, making it easy to reproduce and accessible to the broader research community.
    \item No heavy pretraining: Our approach requires only post-training on a lightweight dataset of several hundred hours, significantly reducing computational demands compared to full pretraining—making it accessible to universities without large-scale computing resources.
\end{itemize}

Based on the current experimental results, there are some reasonable and promising future research topics:
\begin{itemize}
    \item Scaling up data: Our results (Table~\ref{tab:covost-results} and Table~\ref{tab:fleurs-results}) already show the benefits of increasing data size. Further scaling can lead to even greater performance gains.
    \item Pretraining the audio post-LM: Pretraining the audio post-decoder while keeping the language model fixed could further improve alignment between output speech and text. The data could be other kinds of data, not limited to speech translation data.
    \item Exploring extreme low-resource settings: As shown in Table~\ref{tab:covost-results}, our method performs well on Japanese (JA) with only 2 hours of training data. We aim to extend this to other low-resource languages where paired data is scarce, further promoting seamless communication and breaking language barriers.
\end{itemize}

\section{Conclusion}
\label{sec:conclusion}

We present SLM-S2ST, a simple yet effective method to extend speech-aware language models for direct speech-to-speech translation. Built on the Phi4-MM, SLM-S2ST integrates speech understanding and generation within a unified framework using a delayed audio transformer head and a streaming vocoder.
Trained on only 940 hours of open-source CVSS-C data, our model achieves state-of-the-art performance, significantly outperforming prior systems using the same dataset. Scaling the model and training data further boosts performance to match the state-of-the-art model trained on much larger corpora.
Importantly, for our system, ``Ours 4B (CVSS-C)'', from the base model to training data, are open-source and will be much easier to reproduce, making it accessible to the research community. 
This lowers the barrier for academic researchers to contribute to speech language model development, not limited to just the industry.

% \newpage
\bibliographystyle{IEEEtran}
\bibliography{IEEEabrv,refs}

@misc{gpt4o,
      title={Hello GPT-4o}, 
      author={OpenAI},
      year={2024},
      url={https://openai.com/index/hello-gpt-4o/}
}

@article{arora2025landscape,
  title={On the landscape of spoken language models: A comprehensive survey},
  author={Arora, Siddhant and Chang, Kai-Wei and Chien, Chung-Ming and Peng, Yifan and Wu, Haibin and Adi, Yossi and Dupoux, Emmanuel and Lee, Hung-Yi and Livescu, Karen and Watanabe, Shinji},
  journal={arXiv preprint arXiv:2504.08528},
  year={2025}
}

@article{peng2024survey,
  title={A survey on speech large language models},
  author={Peng, Jing and Wang, Yucheng and Xi, Yu and Li, Xu and Zhang, Xizhuo and Yu, Kai},
  journal={arXiv preprint arXiv:2410.18908},
  year={2024}
}

@article{wu2024towards,
  title={Towards audio language modeling-an overview},
  author={Wu, Haibin and Chen, Xuanjun and Lin, Yi-Cheng and Chang, Kai-wei and Chung, Ho-Lam and Liu, Alexander H. and Lee, {Hung-yi}},
  journal={arXiv preprint arXiv:2402.13236},
  year={2024}
}

@article{ji2024wavchat,
  title={Wavchat: A survey of spoken dialogue models},
  author={Ji, Shengpeng and Chen, Yifu and Fang, Minghui and Zuo, Jialong and Lu, Jingyu and Wang, Hanting and Jiang, Ziyue and Zhou, Long and Liu, Shujie and Cheng, Xize and others},
  journal={arXiv preprint arXiv:2411.13577},
  year={2024}
}

@article{defossez2024moshi,
  title={Moshi: a speech-text foundation model for real-time dialogue},
  author={D{\'e}fossez, Alexandre and Mazar{\'e}, Laurent and Orsini, Manu and Royer, Am{\'e}lie and P{\'e}rez, Patrick and J{\'e}gou, Herv{\'e} and Grave, Edouard and Zeghidour, Neil},
  journal={arXiv preprint arXiv:2410.00037},
  year={2024}
}

@article{phi4mini,
      title={Phi-4-Mini Technical Report: Compact yet Powerful Multimodal Language Models via Mixture-of-LoRAs}, 
      author={Microsoft and others},
      year={2025},
      journal="arXiv preprint arXiv:2503.01743",
}

@article{ding2025kimi,
  title={Kimi-audio technical report},
  author={Ding, Ding and Ju, Zeqian and Leng, Yichong and Liu, Songxiang and Liu, Tong and Shang, Zeyu and Shen, Kai and Song, Wei and Tan, Xu and Tang, Heyi and others},
  journal={arXiv preprint arXiv:2504.18425},
  year={2025}
}

@inproceedings{pGSLM,
  title={Text-Free Prosody-Aware Generative Spoken Language Modeling},
  author={Eugene Kharitonov and Ann Lee and Adam Polyak and Yossi Adi and Jade Copet and Kushal Lakhotia and Tu-Anh Nguyen and Morgane Rivière and Abdelrahman Mohamed and Emmanuel Dupoux and Wei-Ning Hsu},
  booktitle={acl},
  year={2022},
}

@article{xu2025qwen2,
  title={Qwen2. 5-omni technical report},
  author={Xu, Jin and Guo, Zhifang and He, Jinzheng and Hu, Hangrui and He, Ting and Bai, Shuai and Chen, Keqin and Wang, Jialin and Fan, Yang and Dang, Kai and others},
  journal={arXiv preprint arXiv:2503.20215},
  year={2025}
}

@article{du2024cosyvoice,
  title={Cosyvoice 2: Scalable streaming speech synthesis with large language models},
  author={Du, Zhihao and Wang, Yuxuan and Chen, Qian and Shi, Xian and Lv, Xiang and Zhao, Tianyu and Gao, Zhifu and Yang, Yexin and Gao, Changfeng and Wang, Hui and others},
  journal={arXiv preprint arXiv:2412.10117},
  year={2024}
}

@inproceedings{radford2023robust,
  title={Robust speech recognition via large-scale weak supervision},
  author={Radford, Alec and Kim, Jong Wook and Xu, Tao and Brockman, Greg and McLeavey, Christine and Sutskever, Ilya},
  booktitle={International conference on machine learning},
  pages={28492--28518},
  year={2023},
  organization={PMLR}
}

@inproceedings{conformer20,
  author       = {Anmol Gulati and
                  James Qin and
                  Chung{-}Cheng Chiu and
                  Niki Parmar and
                  Yu Zhang and
                  Jiahui Yu and
                  Wei Han and
                  Shibo Wang and
                  Zhengdong Zhang and
                  Yonghui Wu and
                  Ruoming Pang},
  title        = {Conformer: Convolution-augmented Transformer for Speech Recognition},
  booktitle    = {21st Annual Conference of the International Speech Communication Association,
                  Interspeech 2020, Virtual Event, Shanghai, China, October 25-29, 2020},
  pages        = {5036--5040},
  publisher    = {{ISCA}},
  year         = {2020},
}

@inproceedings{
mentzer2024finite,
title={Finite Scalar Quantization: {VQ}-{VAE} Made Simple},
author={Fabian Mentzer and David Minnen and Eirikur Agustsson and Michael Tschannen},
booktitle={The Twelfth International Conference on Learning Representations},
year={2024},
url={https://openreview.net/forum?id=8ishA3LxN8}
}

@inproceedings{
lipman2023flow,
title={Flow Matching for Generative Modeling},
author={Yaron Lipman and Ricky T. Q. Chen and Heli Ben-Hamu and Maximilian Nickel and Matthew Le},
booktitle={The Eleventh International Conference on Learning Representations },
year={2023},
url={https://openreview.net/forum?id=PqvMRDCJT9t}
}

@inproceedings{hifigan,
 author = {Kong, Jungil and Kim, Jaehyeon and Bae, Jaekyoung},
 booktitle = {Advances in Neural Information Processing Systems},
 editor = {H. Larochelle and M. Ranzato and R. Hadsell and M.F. Balcan and H. Lin},
 pages = {17022--17033},
 publisher = {Curran Associates, Inc.},
 title = {HiFi-GAN: Generative Adversarial Networks for Efficient and High Fidelity Speech Synthesis},
 url = {https://proceedings.neurips.cc/paper_files/paper/2020/file/c5d736809766d46260d816d8dbc9eb44-Paper.pdf},
 volume = {33},
 year = {2020}
}

@article{hu2022lora,
  title={Lora: Low-rank adaptation of large language models.},
  author={Hu, Edward J and Shen, Yelong and Wallis, Phillip and Allen-Zhu, Zeyuan and Li, Yuanzhi and Wang, Shean and Wang, Lu and Chen, Weizhu and others},
  journal={ICLR},
  volume={1},
  number={2},
  pages={3},
  year={2022}
}

@inproceedings{tang2023salmonn,
title={{SALMONN}: Towards Generic Hearing Abilities for Large Language Models},
author={Changli Tang and Wenyi Yu and Guangzhi Sun and Xianzhao Chen and Tian Tan and Wei Li and Lu Lu and Zejun MA and Chao Zhang},
booktitle={iclr},
year={2024},
}

@article{chu2023qwen,
  title={{Qwen-Audio}: Advancing universal audio understanding via unified large-scale audio-language models},
  author={Chu, Yunfei and Xu, Jin and Zhou, Xiaohuan and Yang, Qian and Zhang, Shiliang and Yan, Zhijie and Zhou, Chang and Zhou, Jingren},
  journal={arXiv preprint arXiv:2311.07919},
  year={2023}
}

@article{gong2023listen,
  title={Listen, think, and understand},
  author={Gong, Yuan and Luo, Hongyin and Liu, Alexander H and Karlinsky, Leonid and Glass, James},
  journal={arXiv preprint arXiv:2305.10790},
  year={2023}
}

@article{ghosh2025audio,
  title={Audio flamingo 2: An audio-language model with long-audio understanding and expert reasoning abilities},
  author={Ghosh, Sreyan and Kong, Zhifeng and Kumar, Sonal and Sakshi, S and Kim, Jaehyeon and Ping, Wei and Valle, Rafael and Manocha, Dinesh and Catanzaro, Bryan},
  journal={arXiv preprint arXiv:2503.03983},
  year={2025}
}

@article{barrault2023seamless,
  title={Seamless: Multilingual Expressive and Streaming Speech Translation},
  author={Barrault, Lo{\"\i}c and Chung, Yu-An and Meglioli, Mariano Coria and Dale, David and Dong, Ning and Duppenthaler, Mark and Duquenne, Paul-Ambroise and Ellis, Brian and Elsahar, Hady and Haaheim, Justin and others},
  journal={arXiv preprint arXiv:2312.05187},
  year={2023}
}

@inproceedings{wang2021covost2,
  title={CoVoST 2 and Massively Multilingual Speech Translation},
  author={Changhan Wang and Anne Wu and Jiatao Gu and Juan Pino},
  booktitle={Proceedings of Interspeech 2021},
  pages={2247--2251},
  year={2021},
  doi={10.21437/Interspeech.2021-2027}
}

@inproceedings{conneau2023fleurs,
  title={Fleurs: Few-shot learning evaluation of universal representations of speech},
  author={Conneau, Alexis and Ma, Min and Khanuja, Simran and Zhang, Yu and Axelrod, Vera and Dalmia, Siddharth and Riesa, Jason and Rivera, Clara and Bapna, Ankur},
  booktitle={2022 IEEE Spoken Language Technology Workshop (SLT)},
  pages={798--805},
  year={2023},
  organization={IEEE}
}

@article{jia2022cvss,
  title={CVSS corpus and massively multilingual speech-to-speech translation},
  author={Jia, Ye and Ramanovich, Michelle Tadmor and Wang, Quan and Zen, Heiga},
  journal={arXiv preprint arXiv:2201.03713},
  year={2022}
}

@inproceedings{lee2022direct,
  title={Direct Speech-to-Speech Translation With Discrete Units},
  author={Lee, Ann and Chen, Peng-Jen and Wang, Changhan and Gu, Jiatao and Popuri, Sravya and Ma, Xutai and Polyak, Adam and Adi, Yossi and He, Qing and Tang, Yun and others},
  booktitle={Proceedings of the 60th Annual Meeting of the Association for Computational Linguistics (Volume 1: Long Papers)},
  pages={3327--3339},
  year={2022}
}

@inproceedings{jia2022translatotron,
  title={Translatotron 2: High-quality direct speech-to-speech translation with voice preservation},
  author={Jia, Ye and Ramanovich, Michelle Tadmor and Remez, Tal and Pomerantz, Roi},
  booktitle={International Conference on Machine Learning},
  pages={10120--10134},
  year={2022},
  organization={PMLR}
}

@article{fang2023daspeech,
  title={Daspeech: Directed acyclic transformer for fast and high-quality speech-to-speech translation},
  author={Fang, Qingkai and Zhou, Yan and Feng, Yang},
  journal={Advances in Neural Information Processing Systems},
  volume={36},
  pages={72604--72623},
  year={2023}
}

@inproceedings{inaguma2023unity,
  title={UnitY: Two-pass Direct Speech-to-speech Translation with Discrete Units},
  author={Inaguma, Hirofumi and Popuri, Sravya and Kulikov, Ilia and Chen, Peng-Jen and Wang, Changhan and Chung, Yu-An and Tang, Yun and Lee, Ann and Watanabe, Shinji and Pino, Juan},
  booktitle={Proceedings of the 61st Annual Meeting of the Association for Computational Linguistics (Volume 1: Long Papers)},
  pages={15655--15680},
  year={2023}
}

@inproceedings{zhang2024streamspeech,
  title={StreamSpeech: Simultaneous Speech-to-Speech Translation with Multi-task Learning},
  author={Zhang, Shaolei and Fang, Qingkai and Guo, Shoutao and Ma, Zhengrui and Zhang, Min and Feng, Yang},
  booktitle={Proceedings of the 62nd Annual Meeting of the Association for Computational Linguistics (Volume 1: Long Papers)},
  pages={8964--8986},
  year={2024}
}

@article{chu2024qwen2,
  title={Qwen2-audio technical report},
  author={Chu, Yunfei and Xu, Jin and Yang, Qian and Wei, Haojie and Wei, Xipin and Guo, Zhifang and Leng, Yichong and Lv, Yuanjun and He, Jinzheng and Lin, Junyang and others},
  journal={arXiv preprint arXiv:2407.10759},
  year={2024}
}

\end{document}